\documentclass[aps,prd,showpacs,twocolumn]{revtex4}
\usepackage{epsfig}
\usepackage[latin1]{inputenc}
\usepackage{amsmath,amssymb}
\usepackage{graphicx}

\begin{document}
\title{
Estimating total momentum at finite distances}

\author{
Emanuel Gallo${}^1$\thanks{Member of CONICET.},
Luis Lehner${}^2$\thanks{
email: lehner@phys.lsu.edu}
and
Osvaldo M. Moreschi${}^1$\thanks{Member of CONICET.}
\thanks{ email: moreschi@fis.uncor.edu} \\
\vspace{3mm}
${}^1$\small FaMAF, Universidad Nacional de C\'{o}rdoba\\
\small Ciudad Universitaria, (5000) C\'{o}rdoba, Argentina. \\
\vspace{3mm}
${}^2$\small Department of Physics \& Astronomy, LSU, \\
\small Baton Rouge, LA 70803, USA.
}

\begin{abstract}
We study the difficulties associated with the evaluation of the total Bondi momentum at finite distances around the central source of a general (asymptotically flat) spacetime.
Since the total momentum is only rigorously defined at future null infinity,
both finite distance and gauge effects must be taken into account
for a correct computation of this quantity.
 Our discussion
is applicable in general contexts but is particularly relevant in numerically
constructed spacetimes for both extracting important physical information and
assessing the accuracy of additional quantities.
\end{abstract}

\maketitle

\section{Introduction}
The possibility of studying highly dynamical, compact
systems by the observation of gravitational waves with the new generation
of gravitational wave detectors is about to become a reality. 
Efforts at the experimental front have allowed the coming on-line of
several highly sensitive detectors which will be followed by even more
refined ones\cite{Frey07,Willke07,Acernese05}. 
Confronting the output with theoretical models will shed light to a number of 
spectacular events which have so far remained out of reach. In fact,  
the analysis of the detectors' output has already provided upper bounds
to several systems, giving insight into the possible 
source or location of a gamma ray burst event\cite{Abbott07},
and constraining important physical parameters in the crab pulsar\cite{Abbott08}.
A detection and subsequent analysis will go further than this by providing
key information of the system that produced it, if accurate models are available to
tie theory with observation.
For binary black hole systems, concentrated efforts are now focusing on providing
effective templates covering from early orbiting stages through the merger and post-merger 
stages and assessing their accuracy\cite{Buonanno07,Ajith07,Damour08,Boyle07}. 
Intimately related to this latter issue is the suitable identification of physical
parameters at each of the different stages (treated by different techniques) and the
extraction of sought after information.
When studying an isolated system in general relativity, there are
several quantities that have strict physical meaning only in the asymptotic
region; for example: gravitational radiation fields, total momentum,
total angular momentum, etc.; which are defined at null infinity.
Numerical approximations however, often are not able to reach
the asymptotic region and so the relevant quantities must either be computed
at finite distances or propagated somehow to infinity for its unambiguous
computation.
In a previous article\cite{Lehner07}, we have analyzed the
difficulties that appear when estimating gravitational radiation at
finite distances and how to remove them. 
In this article, we discuss
the subject of estimating total momentum also at finite distances
from the sources. This is useful to extract further physical information
and to check the numerical solution itself.
In particular we concentrate on the total Bondi momentum which is an important physical quantity. 
For example, it is directly related to the
computation of velocity recoils produced by kicks in collisions of
compact body systems, or in the gravitational radiation recoils
produced in anisotropic collapse of stars, 
etc.\cite{Baker06,Herrmann07,Campanelli07,Campanelli07b,Gonzalez07,Gonzalez07b,Koppitz07,Schnittman08,Lousto08,Baker08}.
The prevalent approach to compute these quantities within a simulation is to
do so through a temporal integration of the Bondi momentum flux. The Bondi momentum
on the other hand needs only be evaluated at the initial and final times for
this purpose. Furthermore a balance law relates the difference between
the Bondi momentum with the integral of the momentum flux, which can be 
employed as a further check of the implementation.

In what follows we review the Bondi momentum calculation at infinity. This
will serve as a motivation for the rest of the work, by addressing the
limitations that finite-distance calculations must address to reduce
the  associated errors when attempting such a task.
The notion of total momentum is normally associated to an integration
on an asymptotic sphere at future null infinity
of an asymptotically flat spacetime.
The integration employs the so called four-translations
of the Bondi-Metzner-Sachs group (BMS), and an appropriate integrand.
A typical expression in terms of standard notation is
\begin{equation}\label{eq:bondimomentum}
P^a =  - \frac{1}{4 \pi} \int_S \hat l^{a} (\Psi_2^0 + \sigma^0 \dot {\bar \sigma}^0) dS^2 ;
\end{equation}
where
$\hat l^a =(1,\sin(\theta) \cos(\phi), \sin(\theta) \sin(\phi), \cos(\theta) ) $ when 
expressed in standard angular coordinates $(\theta,\phi)$ and
all these quantities are calculated in terms of a Bondi coordinate
and tetrad system\footnote{For a review of the conditions for a Bondi coordinate/tetrad
system see \cite{Moreschi86}}. In particular, recall that the Bondi mass
$M$ is the timelike component of this vector; namely
\begin{equation}
M =  - \frac{1}{4 \pi} \int_S (\Psi_2^0 + \sigma^0 \dot {\bar \sigma}^0) dS^2 .
\end{equation}
The Bondi momentum can also be
expressed in terms of the Psi supermomentum\cite{moreschi88}
$\Psi \equiv \Psi_2^0 + \sigma^0 \dot {\bar \sigma}^0
+ \eth^2 \bar\sigma^0$ by
\begin{equation}
P^a =  - \frac{1}{4 \pi} \int_S \hat l^{a} \Psi dS^2 .
\end{equation}
This supermomentum has a couple of useful properties; namely, it is
real, $\bar\Psi = \Psi$, and its time derivative is simply
\begin{equation}
\dot \Psi = \dot \sigma^0 \dot {\bar \sigma}^0.
\end{equation}
Thus, one can see that the time variation of the Bondi momentum,
in terms of time Bondi coordinate, is just
given by
\begin{equation}
\dot P^a =  - \frac{1}{4 \pi } \int_S \hat l^{a} \dot \sigma^0 \dot {\bar \sigma}^0 dS^2 \label{radiate} .
\end{equation}

As we discussed in~\cite{Lehner07}, unless further structure is added
to $3+1$ implementations, these can not access future null infinity
to calculate useful quantities and must resort to do so at finite
distances. This might not only bring about finite size effects but
also difficulties in adopting the appropriate frame --analogous to a Bondi one--
since the advantageous coordinates at the evolution level need not
agree with Bondi-type ones. In~\cite{Lehner07} we examined how these
issues influence the calculation of the radiation and a way to address
them.  We next carry out a similar analysis for the momentum.
Certainly, as the underlying issues are the same, we rely on our
discussion and refer the reader to that work for further details.
For the sake of completeness however, we include here briefly the list 
of basic assumptions required in the calculation of relevant quantities 
at future null infinity (further details are presented in ~\cite{Lehner07}) :
\begin{itemize}
\item Peeling is assumed.
\item Outgoing null hypersurfaces, parameterized by $u$ intersect ${\cal I}^+$
(future null infinity)
defining a sequence of $S^2$ surfaces.
\item Each of these surfaces is conformal to a unit sphere metric; off this surface (into
the spacetime) the departure from it is of lower order. Namely the angular
metric in a neighborhood of ${\cal I}^+$ can be expressed,
as
$g_{AB} = r^2 h_{AB} =  r^2 (q_{AB}/V^2 + c_{AB}/r + O(r^{-2})) $;
with
$V$ a conformal factor, $q_{AB}$ a unit sphere metric, and $r$ a suitably
defined radial distance.
\item A null-tetrad $\{\ell^a,n^a,m^a,\bar m^a\}$ satisfying
$\ell^a n_a = - m^a \bar m_a = 1$ (with all other products being
$0$).  
\item Using standard\cite{Pirani64,Geroch73}
conventions for the Riemann tensor and spinor dyad,
particularly useful scalars obtained from them 
are,
\begin{eqnarray}
\Psi_4 &=&  C_{abcd} n^a \bar m^b n^c \bar m^d ,\\
\Psi_3 &=&  C_{abcd} \ell^a n^b \bar m^c \bar n^d ,\\
\Psi_2 &=&  C_{abcd} \ell^a m^b \bar m^c n^d  ,\label{eq:psi2}\\
\sigma &=& m^a m^b \nabla_a l_b .
\end{eqnarray}
\item A suitable (Bondi type) expansion in terms of $1/r$ with
coordinates chosen such that $V=1$, $g_{ur}^0=1$ and $g_{uA}^0=0$ ($x^A$ labeling
angular coordinates at $u=const$, $r \rightarrow \infty $)
gives rise, in particular, to~\cite{Newman62a,Moreschi87}
\begin{eqnarray}
\Psi_4^0 &=& -\ddot{ \bar \sigma }^0 , \label{eqp4sigma} \\
\Psi_3^0 &=& - \eth \dot{\bar\sigma}^0 , \\
\dot \Psi_2^0 &=&  \eth \Psi_3^0 + \sigma^0 \Psi_4^0 ;
\end{eqnarray}
where the supra-index ``${}^0$'' indicates leading order in an expansion in the radial coordinate
and $\eth$ is the edth operator\cite{Geroch73} of the
unit sphere.
However, if one does not adopt a Bondi frame, the previous equations become much
more complicated.
\end{itemize}

\begin{figure}[h]
\includegraphics{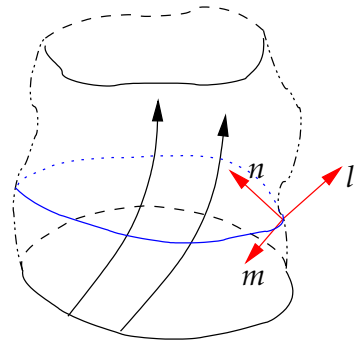}
\caption{\label{f:fig1}. The extraction world-tube $\Gamma$ and the
associated null tetrad and coordinates. The nearly vertical curves
denote the time evolution of the angular coordinates on the
world-tube. The tetrad null vector $\ell$ points outwards, while $n$
points inwards. Complex null vectors $(m,\bar m)$ are tangent to the
spheres contained in the world-tube. }
\end{figure}

Working  with finite size regions one would like to estimate the
asymptotic fields in terms of null tetrads based on a choice of
coordinate system. The nature of the numerical work will suggest the choice
of coordinates --adopted to simplify the simulation-- and consequently
some natural tetrads that can be defined.
These null tetrads however, will not be in general Bondi tetrads.
In order to make these estimations,
an extraction world-tube is assumed to be far
enough to calculate the asymptotic quantities. See figure
\ref{f:fig1}.

The metric is expressed in terms of a null tetrad through
\begin{equation}
g_{ab}=\ell _{a}\;n_{b}+n_{a}\;\ell _{b}-m_{a}\;\overline{m}_{b}-\overline{m}%
_{a}\;m_{b}.
\end{equation}
The null tetrad can be related to a dyad of spinors $(o^A,
\iota^A)$. The relation among the null tetrad vectors and a spinor
dyad is given by $\ell^a \Longleftrightarrow o^A o^{A'}$, $m^a
\Longleftrightarrow o^A \iota^{A'}$, $\bar m^a \Longleftrightarrow
\iota^A o^{A'}$ and $n^a \Longleftrightarrow \iota^A \iota^{A'}$.

Therefore, the transformation of the tetrad frame can be obtained, as
we do here, from the transformation of the dyad frame\footnote{Notice one can
study the transformation of the tetrad frame directly, we employ
the dyad approach as it simplifies the computations involved.}. We will also make use of a
null polar coordinate system $(x^{0},x^{1},x^{2},x^{3})=\left( u,r,%
(\zeta +\overline{\zeta}),\frac{1}{i}(\zeta -\overline{\zeta})\right)$;
where $r$ has the meaning of a radial coordinate
and $(\zeta,\overline{\zeta})$ angular coordinates.

Using the numerical time coordinate $t$,
we construct the  coordinate system
$(u,r,\zeta, \bar\zeta)$ in the following way.
We define $\Gamma$ to be the timelike
surface given by $(r=R=\text{const.})$.
On $\Gamma$, a null tetrad $\{\ell,n,m,\bar m\}$ can be defined in the following way.
Let the null function $u$ be such that on $\Gamma$
one has $u=t$; and $\ell=du$ everywhere. The function $u$ is chosen
such that the future directed vector $\ell$ points outwards with respect to the 2-surfaces
$(t=\text{const.},r=R)$; which are, topologically,
two dimensional spheres.

Then, the complex vectors $m$ and $\bar m$ are defined to be
tangent to the spheres $(u=\text{const.},r=\text{const.})$.
Furthermore, one can choose $m$ to be proportional
to $\frac{\partial}{\partial \zeta}$; and
$\bar m$ to be proportional
to $\frac{\partial}{\partial \bar\zeta}$
in the asymptotic region for large $r$.
The remaining
null vector $n^a$ is set
by requiring the standard normalization conditions.
This is the way in which the coordinate system  $(u,r,\zeta, \bar\zeta)$ is
constructed.

Now, in order to compute the total momentum one must be able to express 
the different Bondi quantities appearing in eq.(\ref{eq:bondimomentum}),
in terms of analogous quantities constructed from the tetrad $\{l^a,n^a,m^a,\bar{m}^a\}$
and coordinate system $(u,r,\zeta, \bar\zeta)$. In the next sections we discuss
how to do so.

\section{Basic structure of asymptotically flat spacetimes}\label{sec:asymptotic}
\subsection{Asymptotic behavior of the tetrad's components}

Let us use the coordinate system
$(u,r,\zeta, \bar\zeta)$.
Keeping $(\zeta=\text{const.})$ and $(\bar\zeta=\text{const.})$
on the null hypersurface $(u=\text{const.})$,
one can see that incrementing $r$ one moves along a null
direction. Since $\ell$ is contained on
the hypersurface $(u=\text{const.})$, one deduces that
$\ell$ is proportional to $\frac{\partial}{\partial r}$.
We will prove below that
\begin{equation}
\left(\ell^a\right) =
\left(\frac{1}{g_{ur}}\frac{\partial}{\partial r}
\right)^a .\label{unoa}
\end{equation}

A natural null tetrad is then completed with,
\begin{equation}
\ell _{a}=\left( du\right)_{a} ,
\label{uno}
\end{equation}
\begin{equation}
m^{a}=\xi ^{i}\left( \frac{\partial }{\partial x^{i}}\right) ^{a} ,
\label{eq:vecm}
\end{equation}
\begin{equation}
\overline{m}^{a}=\overline{\xi}^{i}\left( \frac{\partial }{\partial x^{i}}\right) ^{a} ;
\label{tres}
\end{equation}
\begin{equation}\label{eq:vecn}
n^{a}=\,\left(\frac{\partial}{\partial \,u} \right)^{a}
+ \,U\,\left( \frac{\partial }{\partial \,r}\right)^{a}
+ X^{i}\,\left(\frac{\partial }{\partial \,x^{i}}\right)^{a} ,
\end{equation}
with $i=2,3$  and components $\xi^{i}$, $U$ and $X^{i}$ are:
\begin{equation}
\xi ^{2}=\frac{\xi _{0}^{2}}{r} + O\left(\frac{1}{r^2}\right),
\qquad \xi ^{3}=\frac{\xi_{0}^{3}}{r} + O\left(\frac{1}{r^2}\right) ,
\end{equation}
with
\begin{equation}\label{eq:xileading}
\xi _{0}^{2}=\sqrt{2}P_{0}\;V,\qquad \xi _{0}^{3}=-i\xi _{0}^{2} ;
\end{equation}
where $V=V(u,\zeta,\bar\zeta)$ and
the square of $P_0 = \frac{(1+\zeta \bar\zeta)}{2}$ is the conformal factor
of the unit sphere;
\begin{equation}
U=rU_{00}+U_{0}+\frac{U_{1}}{r} +  O\left(\frac{1}{r^2}\right) ;
\end{equation}
and the other components of the vector $n^a$ have the asymptotic form
\begin{equation}
 X^{2}= X^2_{0} + O\left(\frac{1}{r}\right) ,
\qquad X^{3}= X^3_{0} + O\left(\frac{1}{r}\right)  .
\end{equation}

\subsection{Coordinate and tetrad transformations}
With the objective of computing the total momentum,
one must transform to an asymptotic Bondi
coordinate $(\tilde u, \tilde r, \tilde\zeta,\bar{\tilde\zeta})$
and tetrad frame $(\tilde \ell,\tilde n, \tilde m, \bar{\tilde m})$.

The transformation is of the form
\begin{align}
\tilde u &= \alpha(u,\zeta,\bar\zeta) +
 \frac{\tilde u_1(u,\zeta,\bar\zeta)}{r}  \label{eq:tildeu2}
+ O\left(\frac{1}{r^2}\right) ,\\
\tilde r &= \frac{r}{w(u,\zeta,\bar\zeta)}  \label{eq:tilder2}
+ O\left(r^0\right)  ,\\
\tilde \zeta & = \tilde \zeta_0(u,\zeta,\bar\zeta) + O\left(\frac{1}{r}\right) .\label{eq:tildezeta2}
\end{align}
with $\dot \alpha >0$.

Let us note that in a Bondi system, the angular coordinates
$(\tilde\zeta,\bar{\tilde\zeta})$ can be chosen as the stereographic coordinates of
asymptotic spheres. If one further assumes that $\zeta$ is a
stereographic coordinate of the spheres $(t=\text{const.},
r=\text{const.})$; which are conformally related to the Bondi
coordinates, it is only necessary to consider an angular
transformation of the form\cite{Lehner07}
\begin{equation}
\tilde \zeta = \tilde \zeta_0(u,\zeta) + O\left(\frac{1}{r}\right) .
\label{eq:tildezeta2b}
\end{equation}

The contravariant metric components for the standard Bondi like
coordinate system are given by equations (3.13-18) of
\cite{Moreschi87}; whose inverse is given by equations
(3.19-24) of the same reference.

For the general tetrad the only difference is that now
\begin{equation}
g^{ur}=\frac{1}{g_{ur}} 
\end{equation}
has not necessarily a unit value.

In order to deduce the relations among the dyads let us start from the null
tetrad defined by (\ref{unoa}),
\begin{equation}
\left(\ell^a\right) =
\left(A \frac{\partial}{\partial r}
\right)^a ,
\end{equation}
together with eqns. (\ref{eq:vecm})-(\ref{eq:vecn}). Then the inverse 
metric is given by
\begin{align}
g^{uu} &= 0 ,\\
g^{ur} &= A ,\\
g^{ui} &= 0 ,\\
g^{rr} &= 2 A \, U ,\\
g^{ri} &= A \, X^i ,\\
g^{ij} &= -(\xi^i \bar\xi^j + \bar\xi^i \xi^j) .
\end{align}
While the metric is given by
\begin{align}
g_{ur} &= \frac{1}{A} ,\\
g_{rr} &= 0 ,\\
g_{ri} &= 0 ,\\
g_{uu} &= -2 \frac{U}{A} + X^i X^j g_{ij}  ,\\
g_{ui} &= - g_{ij} X^j ,\\
g_{ij} &= (g^{ij})^{-1} =
  -d \epsilon_{ik} \epsilon_{jl}(\xi^k \bar\xi^l+ \bar\xi^k \xi^l) ;
\end{align}
with $i,j,k,l=2,3$, $d = \det(g_{ij})$,
$\epsilon_{ij} = -\epsilon_{ji}$ and $\epsilon_{23}=1$.
In particular, defining the quantity
\begin{equation}
\lambda = \epsilon_{ij} \xi^i \bar\xi^j ;
\end{equation}
one has that
\begin{equation}
d = \frac{1}{|\lambda|^2} .
\end{equation}

The coordinate transformation induces a tetrad transformation
as described in \cite{Lehner07}. The main resulting equations are:

\begin{equation}\label{eq:dotalfa}
\dot\alpha \frac{1}{w} = g^0_{ur} = \frac{1}{A^0}
;
\end{equation}
which determines a relation among $w$, $\alpha$ and $g^0_{ur}$,
where a superscript `${}^0$' means leading behavior for large $r$.

The function $\tilde \zeta$ must be chosen\cite{Lehner07}
such that
\begin{equation}
\dot{\bar{\tilde \zeta}} + X^{0\,\bar \zeta} \;\bar{\tilde \zeta}_{\bar \zeta}
= 0 ; \label{eq:condzeta0}
\end{equation}
so that angular coordinates remain constant along
the generators of future null infinity. In particular if
to leading order $g_{ui}=0$ ($X^i=0$), one could adopt initial
conditions so that $\tilde \zeta = \zeta$. In the general
case however, the coordinates will ``shift'' around the world
tube and the explicit transformation must be taken into account 
as described in \cite{Bishop97}.

The transformation of the vector $n$ is given by
\begin{equation}
\begin{split}
\tilde n &=
 \frac{\partial}{\partial \tilde u}  + O\left( \frac{1}{r}\right) \\
&=
\frac{1}{\dot \alpha
- \alpha_\zeta \frac{\dot{\tilde {\zeta}}_0}{\tilde {\zeta_0}_\zeta}
- \alpha_{\bar\zeta} \frac{\dot{\bar{\tilde \zeta}}_0}{{\bar{ \tilde \zeta}_0}_{\bar\zeta}}}
\;
n + O\left( \frac{1}{r}\right)
.
\end{split}\label{eq:trsn}
\end{equation}

Similarly, the asymptotic behavior of the vector $\tilde m$ is given by
\begin{equation}
\begin{split}
\tilde m &=
\frac{\sqrt{2}\tilde P}{\tilde r}  \frac{\partial}{\partial \tilde \zeta}
  + O\left( \frac{1}{r^2}\right) \\
&=
-
\frac{\sqrt{2}\tilde P_0 \tilde V\, w }{ r}
\frac{\alpha_\zeta}
{n^0(\alpha)\;\tilde\zeta_\zeta}
\,
n^0
+
\frac{\tilde P_0 \tilde V \, w}{P_0 V \tilde\zeta_\zeta}
m
+ O\left( \frac{1}{r^2}\right) ;
\end{split}\label{eq:trsm}
\end{equation}
where $n^0$ is the vector $n$ evaluated at future null infinity;
\begin{equation}\label{eq:ncero}
n^0=
\frac{\partial}{\partial u}
+X^{0\,\zeta} \frac{\partial}{\partial \zeta}
+X^{0\,\bar \zeta} \frac{\partial}{\partial \bar\zeta}
;
\end{equation}
and therefore an operator acting on functions.

Since the angular part of the metric expressed in terms of the new null
 tetrad must coincide with the angular part of the metric
expressed in terms of the original null tetrad, it follows that
\begin{equation}\label{eq:tildeVa1}
\frac{P_0^2 V^2}{w^2 \tilde P_0^2 \tilde V^2}
\;\tilde\zeta_\zeta \bar{\tilde\zeta}_{\bar\zeta}
=
1
.
\end{equation}
So that
\begin{equation}\label{eq:tildeVa}
\frac{w \tilde P_0 \tilde V}{P_0 V}
=
\sqrt{\tilde\zeta_\zeta \bar{\tilde\zeta}_{\bar\zeta}}
\; ;
\end{equation}
and therefore the factor of $m$ is
\begin{equation}\label{eq:tildeVa3}
\frac{w \tilde P_0 \tilde V}{P_0 V \tilde\zeta_\zeta}
=
\sqrt{\frac{ \bar{\tilde\zeta}_{\bar\zeta}}{\tilde\zeta_\zeta}}
\; .
\end{equation}

Then, in particular, for a Bondi system one has
\begin{equation}\label{eq:tildeV2}
1=\tilde V
=
\frac{P_0 V
\sqrt{\tilde\zeta_\zeta \bar{\tilde\zeta}_{\bar\zeta} }
}
{\tilde P_0 w }
=
\frac{P_0 V g^0_{ur}
\sqrt{\tilde\zeta_\zeta \bar{\tilde\zeta}_{\bar\zeta} }
}
{\tilde P_0 \dot \alpha}
;
\end{equation}
where we are using that the tilde system is of Bondi type.
Equation (\ref{eq:tildeV2}) determines $\alpha$. 






From the transformation expressed in equations
(\ref{eq:trsn}) and (\ref{eq:trsm}) and the requirement that
the spinor metric $\epsilon_{AB}$ be invariant,
one can deduce the transformation for the spinor dyad
\begin{equation}\label{eq:iotatilde1}
\tilde \iota^A = \frac{1}
{\sqrt
{n^0(\alpha)}
}
\left({\frac{ \tilde\zeta_\zeta}{ \bar{\tilde\zeta}_{\bar\zeta} } }\right)^{\frac{1}{4}}
\, \iota^A \; ;
\end{equation}
and
%
\begin{eqnarray}
\label{eq:omitilde1}
\tilde o^A &=&
{\sqrt
{n^0(\alpha)}
}
\left({\frac{ \tilde\zeta_\zeta}{ \bar{\tilde\zeta}_{\bar\zeta} } }\right)^{\frac{1}{4}}
\left[
\left(
\frac{\tilde P_0 \tilde V \, w}{P_0 V \tilde\zeta_\zeta}
\right)
\, o^A \right. \nonumber \\
 & & -
\left.
\left(
\frac{\sqrt{2}\tilde P_0 \tilde V\, w }{ r} \;
\frac{\alpha_\zeta}{n^0(\alpha) \tilde\zeta_\zeta}
\right)
\, \iota^A
\right] \nonumber \\
&=&
\sqrt{n^0(\alpha)}
\left({\frac{ \tilde\zeta_\zeta}{ \bar{\tilde\zeta}_{\bar\zeta} } }\right)^{\frac{1}{4}}
\frac{\tilde P_0 \tilde V \, w}{P_0 V \tilde\zeta_\zeta}
\left(
 o^A
-
\frac{1}{r} \;
\frac{\eth_V \alpha}{n^0(\alpha)}
\, \iota^A
\right) \nonumber \\
&=&
\sqrt{n^0(\alpha)}
\left({\frac{ \bar{\tilde\zeta}_{\bar\zeta} }{ \tilde\zeta_\zeta} }\right)^{\frac{1}{4}}
\left(
 o^A
-
\frac{1}{r} \;
\frac{\eth_V \alpha}{n^0(\alpha)}
\, \iota^A
\right)
\; .
\end{eqnarray}

%
%

The regular dyad at future null infinity is then given by
\begin{equation}\label{eq:regulariota}
\hat{\tilde \iota}^A = \tilde \iota^A =
\frac{1}
{\sqrt
{n^0(\alpha)}
}
\left({\frac{ \tilde\zeta_\zeta}{ \bar{\tilde\zeta}_{\bar\zeta} } }\right)^{\frac{1}{4}}
\,
\hat \iota^A ;
\end{equation}
and
\begin{eqnarray}\label{eq:regularomicrom}
\hat{\tilde o}^A &=& \tilde\Omega^{-1}\,  \tilde o^A \nonumber \\
&=&
\frac{r}{w}
\sqrt{n^0(\alpha)}
\left({\frac{ \bar{\tilde\zeta}_{\bar\zeta} }{ \tilde\zeta_\zeta} }\right)^{\frac{1}{4}}
\left(
 o^A
-
\frac{1}{r} \;
\frac{\eth_V \alpha}{n^0(\alpha)}
\, \iota^A
\right)
 \nonumber \\
&=&
\frac{1}{w}
\sqrt{n^0(\alpha)}
\left({\frac{ \bar{\tilde\zeta}_{\bar\zeta} }{ \tilde\zeta_\zeta} }\right)^{\frac{1}{4}}
\left(
\hat  o^A
-
\frac{\eth_V \alpha}{n^0(\alpha)}
\, \hat \iota^A
\right) \nonumber \\
&=&
B
\left(
\hat  o^A
+C
\, \hat \iota^A
\right) \; ;
\end{eqnarray}
where
\begin{equation}
B =
\frac{1}{w}
\sqrt{n^0(\alpha)}
\left({\frac{ \bar{\tilde\zeta}_{\bar\zeta} }{ \tilde\zeta_\zeta} }\right)^{\frac{1}{4}}
=
\frac{g^0_{ur}}{\dot\alpha}
\sqrt{n^0(\alpha)}
\left({\frac{ \bar{\tilde\zeta}_{\bar\zeta} }{ \tilde\zeta_\zeta} }\right)^{\frac{1}{4}}
\end{equation}
and
\begin{equation}
C = -
\frac{\eth_V \alpha}{n^0(\alpha)} .
\end{equation}

Notice that one could generalize (\ref{eq:iotatilde1}) and (\ref{eq:omitilde1})
to include higher order terms in an asymptotic expansion of the
form
\begin{equation}\label{eq:iotatilde2}
\tilde \iota^A = \frac{1}
{\sqrt
{n^0(\alpha)}
}
\left({\frac{ \tilde\zeta_\zeta}{ \bar{\tilde\zeta}_{\bar\zeta} } }\right)^{\frac{1}{4}}
\left(
 \iota^A
+h o^A
\right)
;
\end{equation}
and
\begin{equation}
\label{eq:omitilde2}
\tilde o^A =
w B
\left(
 o^A
(1+\delta)
+\Omega C
(1+\gamma)
\, \iota^A
\right)
;
\end{equation}
however its contribution to the (spinorial) metric
is of higher order as shown by the following expression
\begin{equation}\label{eq:epsilontilde}
\tilde\epsilon^{AB} =
\left[
(1+\delta)
-
\Omega h C (1+\gamma)
\right] \epsilon^{AB} ;
\end{equation}
which in turn implies,
\begin{equation}
\delta = \Omega h C (1+\gamma) .
\end{equation}

\section{Calculation of the Bondi momentum}\label{sec:bondimom}
In order to compute the Bondi momentum, as expressed in
eq. (\ref{eq:bondimomentum}), one needs to evaluate
the Bondi quantities $\tilde\Psi_2^0$,
$\tilde\sigma^0$ and $\dot {\bar {\tilde{\sigma}}}^0$
in terms of the quantities obtained in the
more readily accessible quantities to numerical implementations.

\subsection{Transformation of $\Psi_2^0$}
We can now easily calculate the component $\Psi_2$ of the Weyl tensor,
in leading orders,
with respect to the new null tetrad, obtaining
\begin{equation}\label{eq:forma1psi2}
\begin{split}
\tilde \Psi_2^0 &=
\tilde{\Omega}^{-1} \Psi_{ABCD}
\hat{\tilde o}^A
\hat{\tilde o}^B
\hat{\tilde \iota}^C
\hat{\tilde \iota}^D \\
&=
\frac{B^2}{w \, n^0(\alpha)}
\left({\frac{ \tilde\zeta_\zeta}{ \bar{\tilde\zeta}_{\bar\zeta} } }\right)^{\frac{1}{2}}
\left( \Psi_2^0 -
2\frac{\eth_V\, \alpha}{n^0(\alpha)} \Psi_3^0 + \frac{(\eth_V
\alpha)^2}{n^0(\alpha)^2}\, \Psi_4^0 \right) \\
&=
\left(
\frac{g^0_{ur}}{\dot \alpha}
\right)^3
\left( \Psi_2^0 -
2\frac{\eth_V\, \alpha}{n^0(\alpha)} \Psi_3^0 + \frac{(\eth_V
\alpha)^2}{n^0(\alpha)^2}\, \Psi_4^0 \right)
.
\end{split}
\end{equation}

Note that the asymptotic transformation of the dyad $\left\{o^A,\iota^A\right\}$ can be
re-interpreted as a combination of the null rotations
of types II (rotation around $n^a$) and  III (boost/spin) in the standard Newman Penrose 
formalism (see Appendix A). If one makes  use of these rotations to compute the 
transformation of the complete scalar $\Psi_2$ an expression similar
to eq.~(\ref{eq:forma1psi2}) is obtained but {\it without} the factor 
$\left(g^0_{ur}\, {\dot \alpha^{-1}}\right)^3$. 
This factor is present in eq.~(\ref{eq:forma1psi2}) since we are computing the 
leading order behaviors, which due to the assumed peeling property obey,
\begin{eqnarray}
\Psi_2&=&\frac{\Psi_2^0}{r^3}+O\left(\frac{1}{r^4}\right);\\
\tilde\Psi_2&=&\frac{\tilde\Psi_2^0}{\tilde r^3}+O\left(\frac{1}{\tilde r^4}\right)\, .
\end{eqnarray}
Then, taking into account the relation
\begin{equation}
 \tilde r =\left(\frac{g^0_{ur}}{\dot \alpha}\right) r+O\left(r^0\right);
\end{equation}
if follows that the aforementioned factor will be present in the transformation of $\Psi_2^0$.

\subsection{Transformation of $\tilde\sigma^0$ and $\dot {\bar {\tilde{\sigma}}}^0$ }

Let us recall that the shear is defined by
\begin{equation}
\sigma = o^A \bar\iota^{A'}\, o^B \nabla_{AA'} o_B
= \frac{\sigma_0}{r^2} +O\left(\frac{1}{r^3}\right) .
\end{equation}

Only the leading order transformation of the dyad is needed to
calculate the transformation of the leading order behavior of the
shear.

Then, one has
\begin{equation}
\begin{split}
\tilde\sigma
=
(w B)^2
&
\left({\frac{ \bar{\tilde\zeta}_{\bar\zeta} }{ \tilde\zeta_\zeta} }\right)^{\frac{1}{2}}
\left(
o^A + C \, \Omega \; \iota^A
\right)
\iota^{A'}
\left(
o^B + C \, \Omega \; \iota^B
\right)\\
&
\nabla_{AA'}
\left(
o_B + C \, \Omega \; \iota_B
\right) \\
=
(w B)^2
&
\left({\frac{ \bar{\tilde\zeta}_{\bar\zeta} }{ \tilde\zeta_\zeta} }\right)^{\frac{1}{2}}
\left[
\sigma
-\Omega \, \hat m(\Omega C)
- \Omega C \hat n(\Omega C)
\right. \\
& \Omega C \tau
+2 \Omega C \left( \beta
- \Omega C \epsilon'  \right) \\
&
+
\left.
\Omega^2 C^2
\left(
-\rho'  -\Omega C \kappa'
\right)+ O(\Omega^3)
\right]
;
\end{split}
\end{equation}
where
$\kappa=0$.

The details of the calculations leading to
$\tilde\sigma_0$
are given in the Appendix B. One finds
\begin{equation}\label{eq:sigmatransformado}
\tilde \sigma_0 = B^2 
\left({\frac{ \bar{\tilde\zeta}_{\bar\zeta} }{ \tilde\zeta_\zeta} }\right)^{\frac{1}{2}}
\left[ \sigma_0 - \eth_V C + 2 \frac{C}{A} \,
\eth_V A
- C \, n^0(C)
\right] ;
\end{equation}
and where $C$ is recognized as a quantity of boost-spin weight $\{ p,q\}$ type
$\{2,0\}$ \cite{Geroch73}.

In the calculation of the Bondi momentum, one also needs
the time derivative of the shear that is given by
\begin{equation}\label{eq:sigmapuntotransformado}
\begin{split}
\dot{\tilde \sigma}_0
&\equiv
\frac{\partial \tilde \sigma_0}{\partial \tilde u} \\ 
&=
\frac{1}{n^0(\alpha)} 
n^0\Bigg(
B^2
\left({\frac{ \bar{\tilde\zeta}_{\bar\zeta} }{ \tilde\zeta_\zeta} }\right)^{\frac{1}{2}}
\bigr.\\
&\quad
\bigl.
\left[
\sigma_0
-
\eth_V C
+ 2 \frac{C}{A} \, \eth_V A
- C \, n^0(C)
\right]
\Biggr)
\; .
\end{split}
\end{equation}


Then, we see that by expressing the Bondi scalars
$\tilde\Psi^0_2$, $\tilde\sigma_0$ and $\dot{\tilde\sigma}_0$ in
terms of $\Psi^0_2$, $\sigma_0$ and $\dot\sigma_0$, we get non-trivial
factors and terms containing 
$\Psi^0_3$, $\Psi^0_4$, $g^0_{ur}$, $\tilde\zeta$, $\alpha$,
$V$ and some of their derivatives. This is an analogous situation 
as that observed in
\cite{Lehner07} for the radiation field $\Psi^0_4$ where these correcting
terms will have in general non-trivial angular dependence which would
affect relevant quantities.
These would be important, for example, in computations of kicks
observed in collisions of different masses or spinning compact
objects where the merged black hole moves through the computational
grid and, consequently, the extraction sphere will be non-centered
introducing extra multipolar structure. 

For completeness of the discussion we point out how to
calculate the auxiliary functions: the conformal function $V$,
was explained in our previous article \cite{Lehner07}, the angular
transformation $\tilde\zeta$ is determined by the angular components
of the vector $n$, while $\alpha$ is determined from the knowledge
of $g_{ur}$, $\tilde\zeta$ and $V$.

\section{Examples}
In what follows we discuss two examples to illustrate the  
effects brought by non-adapted tetrads in the calculation
of the momentum. The first example shows how the computation
can be affected already at future null infinity and without
even introducing a 3+1 description of the spacetime. The second
example illustrates how coordinates not adapted to a Bondi
system in a 3+1 description can give rise to results with spurious
gauge dependence.

\subsection{Robinson-Trautman metrics}
As a first example of applications of our expression for the Bondi
momentum, let us study the Robinson-Trautman metrics. These metrics
are vacuum solutions of Einstein's equations containing a congruence
of diverging null geodesics, with vanishing shear and
twist\cite{Robinson62}. They can be written as\cite{Frittelli92}

\begin{equation} \label{pr1}
 ds^{2} =\left( -2Hr+K-2\frac{M(u)}{r} \right) du^{2}
+2\;du\;dr-\frac{r^{2} }{P^{2} } d\zeta \;d\bar{\zeta }
;
\end{equation}
where $P=P(u,\zeta ,\bar{\zeta } )$, $H=\frac{\dot{P} }{P}$,
$K=\Delta \ln P$, a doted quantity denotes its time derivative and
$\Delta$ is the two-dimensional Laplacian for the two-surfaces
$u=$constant, $r=$constant with line element
\[
dS^{2} =\frac{1}{P^{2} } \;d\zeta \;d\bar{\zeta }.
\]

We can describe this line element in terms of the line element of
the unit sphere; this is done by expressing $P$ in terms of
 $P=V(u,\zeta ,\bar{\zeta } )P_{0} (\zeta ,\bar{\zeta } )$, with
 $P_{0} $ the value of $P$ for the unit sphere.
 The function $V(u,\zeta,\bar{\zeta } )$, must satisfy a fourth order parabolic differential equation
 \begin{equation} \label{pr2}
-3\;M\;\dot{V} =V^{4} \;\eth^{2}\bar \eth^2 V-V^3\eth^{2} V  \bar
\eth^2 V ;
\end{equation}
where we fixed the freedom of redefining $u$ in such a way that
$M(u)$ is actually a constant. This equation is known as the
Robinson-Trautman (RT) equation. If $l$ denotes the vector field
that generates the null congruence, then  $l_a=du$, and
$l^a=\left(\frac{\partial}{\partial r}\right)^a$. By completing the
tetrad in the usual way with the three null vectors,
\begin{eqnarray}
n^a&=&\left(\frac{\partial}{\partial u}\right)^a
+\left( Hr - \frac{K}{2} + \frac{M}{r} \right)
\left(\frac{\partial}{\partial r}\right)^a,\\
m^a&=&\frac{\sqrt{2}P}{r}\left(\frac{\partial}{\partial\zeta}\right)^a,\\
\bar{m}^a&=&\frac{\sqrt{2}P}{r}\left(\frac{\partial}{\partial\bar{\zeta}}\right)^a,
\end{eqnarray}
we have a frame adapted to the geometry. We call this
coordinate/tetrad system RT frame.

The relation between the RT coordinate system and a Bondi system
$\{\tilde{u},\tilde{r},\tilde{\zeta},\tilde{\zeta}\}$
 will be of the form given by eqs. (\ref{eq:tildeu2}), (\ref{eq:tilder2}) and (\ref{eq:tildezeta2}).

Now, let us compute the Bondi linear momentum
\begin{equation} \label{pr7}
P^{a} =-\frac{1}{4\pi } \int \left( \tilde{\Psi} _{2}^{0}
+\tilde{\sigma} ^{0} \frac{\partial \tilde{\bar{\sigma} } ^{0}
}{\partial \tilde{u}}  \right) \hat{l}^{a} \;d\tilde{S}^{2}
;
\end{equation}
in a given $u=u_o=\text{const}$, RT section, and let us write this
$P^a$ in terms of an RT frame. The relation between
Robinson-Trautman and Bondi quantities are given by
eqs.(\ref{eq:forma1psi2}), (\ref{eq:sigmatransformado}) and
(\ref{eq:sigmapuntotransformado}). But in order to simplify the
expressions we will use the freedom of BMS group in such a way that
the ``origin" of the new Bondi system $\tilde{u}=0$ coincides with
the RT section $u_0$, where we wish to compute the momentum. That
is, the relation between the Bondi and RT coordinates is such that
$\alpha(u_0,\zeta,\bar{\zeta})=0$. This implies in particular, that
$\eth_V \alpha(u_0,\zeta,\bar{\zeta})=\bar{\eth}_V
\alpha(u_0,\zeta,\bar{\zeta})=0.$

For Robinson-Trautman metrics we have:
\begin{eqnarray}\label{cantidades RT}
\Psi_{2RT}^0&=&-M,\\
\sigma_{RT}&=&\sigma^0 = 0,\\
g_{ur}&=&1,\\
\dot{\alpha}&=&V.
\end{eqnarray}
then in terms of these quantities, and using the
eqs.(\ref{eq:forma1psi2}), (\ref{eq:sigmatransformado}) and
(\ref{eq:sigmapuntotransformado}), we have that the Bondi
momentum reads,
\begin{equation} \label{pr7b}
P^{a} =\frac{1}{4\pi } \int \frac{M}{V^{3} }  \hat{l} ^{a} \;dS^{2}.
\end{equation}
Here it is important to emphasize that the absence of the $V^{-3}$
correcting factor would imply an erroneous constant mass result.

Note additionally, that if we wish to compute the Bondi momentum at two
different RT times
using eq.(\ref{pr7b}), we will obtain two Bondi momenta calculated
with respect to two different Bondi systems which are related by some 
BMS transformation.
In order to compare the second momentum with the first, one should
express it with respect to the first Bondi system.

\subsection{ Vaidya Solution}
As a second example let us study the Vaidya metric which represents
a spherically symmetric solution radiating a null fluid. The usual
representation of this metric is
\begin{equation}\label{vaydia}
ds^2= \left(1-2\frac {M(\tilde{u})}{\tilde{r}
}\right)d{\tilde{u}^2}+2d\tilde{u}d\tilde{r}-\tilde{r}^2d\tilde{\Omega}^2
\end{equation}
where
\begin{equation}
d\tilde{\Omega}^2=d\tilde{\theta}^2+\sin^2\tilde{\theta}
d\tilde{\phi}^2,
\end{equation}
and $\{\tilde{u},\tilde{r},\tilde{\theta},\tilde{\phi}\}$ define a
null polar coordinate system. The mass of this spacetime at a given time
$\tilde u$ is given by $M(\tilde{u})$ which decreases with time.
Notice that the coordinate system used to express the line element
is naturally of Bondi type with an associated tetrad given by
\begin{eqnarray}
\tilde{l}^a&=&\partial^a_{\tilde{r}},\\
\tilde{n}^a&=&{\partial^a_{\tilde{u}}}-\frac{1}{2}\left(1-\frac{2M(\tilde{u})}{\tilde{r}}\right){\partial^a_{\tilde{r}}}\\
\tilde{m}^a&=&\frac{1}{\sqrt{2}\tilde{r}}\left(\partial^a_{\tilde{\theta}}+\frac{i}{\sin\tilde{\theta}}\partial^a_{\tilde{\phi}}\right),\\
\bar{\tilde{m}}^a&=&\frac{1}{\sqrt{2}\tilde{r}}\left(\partial^a_{\tilde{\theta}}-\frac{i}{\sin\tilde{\theta}}\partial^a_{\tilde{\phi}}\right).
\end{eqnarray}
Thus leading order behavior of $\tilde{n}^a$ is
\begin{equation}
\tilde{n}^a=\partial^a_{\tilde{u}}+O\left(\frac{1}{\tilde{r}^2}\right);
\end{equation}
in terms of an asymptotic regular frame at future null infinity.

We can now compute some geometrical quantities pertinent to the present
discussion with this tetrad. In particular we obtain for the Weyl's
scalars and the shear are
\begin{eqnarray}
\tilde{\psi}_{1}&=&\tilde{\psi}_{3}=\tilde{\psi}_{4}=\tilde{\sigma}=0,\\
\tilde{\psi}_{2}&=&-\frac{M(\tilde{u})}{\tilde{r}^3} ;
\end{eqnarray}
thus, in particular, the only non-vanishing component of
the leading order behavior of Weyl's scalar is
$\tilde{\psi}^0_{2}=-M(\tilde{u})$, and the null congruence
determined by $\tilde{l}^a$ is a free shear congruence
($\tilde{\sigma}=0$). If we compute the Bondi momentum to this
metric we obtain the well know result
\begin{equation}
P^{\underline{a}}=
-\frac{1}{4\pi}\int(\tilde{\psi}^0_{2}+\tilde{\sigma}^0\dot{\bar{\tilde{\sigma}}}^0)
\hat{l}^{\underline{a}}
d\tilde{S}^2=(M(\tilde{u}),0,0,0);
\end{equation}
where ${\underline{a}}$ is a numerical index.
This relation shows that $M(\tilde{u})$ is the Bondi mass, and that this
particular Bondi system is at rest with the source (the spatial part of
the Bondi momentum is zero).

So far the discussion has been centered around a null coordinate
system. However, most numerical simulations adopt a natural $3+1$
coordinate system and it is with respect to this system that
relevant quantities are calculated. In order to make contact with
such task, we must re-express the line element in terms of 
a $3+1$ coordinate system through some given transformation 
$(\tilde u, \tilde r, \tilde \theta, \tilde \phi) \rightarrow (T,R,\Theta,\Phi)$.
Naturally different transformations
can be envisaged and in particular ones that will be consistent with a
Bondi system. However, our goal here is to illustrate the issues discussed in the
previous sections in a simple scenario where the coordinates do not necessarily
conform to Bondi ones. We therefore introduce the following transformation,

\begin{eqnarray}
T&=&  \Delta(\tilde u )+\frac{\tilde r}{\Delta^{(1)}(\tilde u )},\\
R&=&\frac{\tilde r}{\Delta^{(1)}(\tilde u )},\\
\Theta&=&\tilde\theta,\\
\Phi&=&\tilde\phi,
\end{eqnarray}
where $\Delta^{(1)}=\frac{d\Delta}{d\tilde u}$ and in general  we
will have $\Delta^{(n)}=\frac{d^n \Delta}{d\tilde u^n}$.
Incidentally, let us note that $\Delta$ is playing the role of the inverse
of the transformation $\alpha$ presented previously.
Therefore we have,
\begin{eqnarray}
d\tilde u &=& \frac{1}{\Delta^{(1)}}(dT - dR)  \; , \\
d\tilde r &=& \frac{\Delta^{(2)} R}{\Delta^{(1)}} dT + \left( \Delta^{(1)} - \frac{\Delta^{(2)} R}{\Delta^{(1)}} \right) dR \;  \\
d\tilde \theta &=& d\Theta \; , \\
d\tilde \phi &=& d\Phi \; .
\end{eqnarray}

With this transformation, we can construct a ``standard"
null tetrad in the usual way as in numerical efforts,
i.e., from a unit vector $N^a$ orthogonal to the surface
$T=\text{const.}$, and three vectors
$\{R^a,\partial^a_\Theta,\partial^a_\Phi\}$ obtained on the sphere
defined by  the $T=\text{const.}$, $R=\text{const.}$ by Gram-Schmidt procedure. The
result is,

\begin{eqnarray}
l'^a&=&\frac{1}{\sqrt{2}}(N^a+R^a)=\frac{1}{\sqrt{2}N}\left[\partial_{T}
+\partial_{R}\right],\\
n'^a&=&\frac{1}{\sqrt{2}}(N^a-R^a)\\\nonumber
&=&\frac{1}{\sqrt{2}N}\left[\partial_{T}-(2N^2-1)\partial_{R}\right],\\
m'^a&=&\frac{1}{\sqrt{2} \Delta^{(1)} R}\left(\partial^a_{\Theta}
+\frac{i}{\sin\Theta}\partial^a_{\Phi}\right),\\
\bar{m}'^a&=&\frac{1}{\sqrt{2} \Delta^{(1)} R}\left(\partial^a_{\Theta}
-\frac{i}{\sin\Theta}\partial^a_{\Phi}\right) ;
\end{eqnarray}
where
\begin{equation}
N=
\sqrt{
2-\frac{1}{(\Delta^{(1)})^2}
\left(1-2\frac{M(\tilde u)}{\tilde r}
+\frac{2 \Delta^{(2)}}{\Delta^{(1)}}\tilde r\right)
}
;
\end{equation}
and where $\tilde r$ and $\tilde u$ must be thought of as functions
of $\{T,R\}$.
If we compute relevant scalar quantities with this tetrad, the only non
vanishing scalar is $\psi'^0_2$ which is,
\begin{equation}
\psi'^0_2=-\frac{M(\tilde u)}{(\Delta^{(1)})^3}
;
\end{equation}
where one finds an additional non-trivial factor which is completely coordinate dependent.
In order to obtain the correct expression we need to take into
account the corrective factors that were discussed in section \ref{sec:bondimom}.
For this, assuming one starts with the $3+1$ version of the Vaidya metric induced
by the coordinates $(T,R,\Theta,\Phi)$, and following our discussion we
need first a null coordinate/tetrad system
naturally induced by these coordinates, namely
\begin{eqnarray}
u&=&T-R,\\
r&=&R,\\
\theta&=&\Theta,\\
\phi&=&\Phi.
\end{eqnarray}
We can now construct the `naturally' induced tetrad $\{l^a,n^a,m^a,\bar{m}^a\}$
to leading order and compare this with the Bondi tetrad
$\{\tilde{l}^a,\tilde{n}^a,\tilde{m}^a ,\bar{\tilde{m}}^a\}$:
\begin{eqnarray}
l_a&=&(du)_a,\;\;l^a=\partial^a_r=\Delta^{(1)}\tilde{l}^a\label{afinl}\\
n^a&=&\partial^a_u+O\left(\frac{1}{r}\right)=\frac{1}{\Delta^{(1)}}\tilde{n}^a+ O\left(\frac{1}{r}\right),\label{ninducido}\\
m^a&=&\frac{1}{\sqrt{2}\Delta^{(1)}r}\left(\partial^a_{\theta}+\frac{i}{\sin\theta}\partial^a_{\phi}\right)=\tilde{m}^a,\\
\bar{m}^a&=&\frac{1}{\sqrt{2}\Delta^{(1)}r}\left(\partial^a_{\theta}-\frac{i}{\sin\theta}\partial^a_{\phi}\right)=\bar{\tilde{m}}^a.
\end{eqnarray}
Then, from eqs.(\ref{afinl})  and (\ref{ninducido}), we see that
\begin{eqnarray}
g_{ur}=1,\;\; \dot\alpha&=&\frac{1}{\Delta^{(1)}}.
\end{eqnarray}
If we compute $\psi^0_2$ with this tetrad we obtain,
\begin{equation}
 \psi^0_2=\psi'^0_2=-\frac{M(\tilde u)}{(\Delta^{(1)})^3}
.
\end{equation}
Then, employing equation (\ref{eq:forma1psi2}), we get
\begin{equation}
\tilde{\psi}^0_2
=\left(\frac{1}{\dot\alpha}\right)^3\psi^0_2
=-M(\tilde u);
\end{equation}
which yields the correct result.

As a final comment, let us note that $\psi_2=\psi'_2$. 
This is
due to the fact that the relation between the two tetrads is
\begin{eqnarray}
l'^a &=& \frac{1}{\sqrt{2}N}l^a ,\\
n'^a &=& \sqrt{2}N n^a ,\\
m'^a &=& m^a,\\
\bar{m}'^a&=&\bar m^a ;
\end{eqnarray}
and therefore the $\sqrt{2} N$ factors cancel.

\section{Final comments}
The calculation of total momentum at finite distances can be employed
to extract valuable information from a numerically obtained spacetime both
to obtain relevant physical quantities and check related results obtained
in the simulation.
In particular, the final black hole recoil velocity can be readily obtained and
the total energy radiated during the length of the simulation.
Standard calculations of kicks in numerical relativity
have relied on integrating the flux of energy over a period of 
interest;
alternatively, it could also be
computed simply by the difference between initial and final space-time momenta; provided
that gauge issues are addressed so that they do not negatively influence either 
the calculation or the comparison
between the results of the two methods.
Our work provides a way of dealing with possible problems and sets up a common
framework, consistent with Bondi's construction, to remove ambiguities
from the calculation and
facilitate the comparison of results obtained by different numerical implementations.

It is probably worthwhile to note that we have been concerned with the notion of
total momentum at future null infinity, as one is interested in
the relation between total momentum and outgoing radiation which can
carry momentum. For this reason our approach has been to
imagine that topological closed 2-surfaces at finite but large distances
can be understood as limiting spheres in the vicinity of future null infinity.
However, once one has a finite closed 2-surface, one can wonder whether to
 regard it
as a limiting sphere in a vicinity of past null infinity instead.
With this point of view, one should relate the difference in the
total momentum at two different finite spheres with the incoming
radiation; which in this case will be characterized by the
leading order behavior, at constant advanced time, of $\Psi_0$.
Consequently the notion of total momentum at finite spheres will give different
quantities whether one regards it as limiting spheres to future null
infinity or to past null infinity.
This brings the issue of whether one should consider both contributions
when dealing with finite spheres. 
The answer is negative if one wants to describe what 
is observable at very long distances from the sources; since, for all
practical purposes, an observer will be situated at future null
infinity. Therefore, the notion of total momentum and its flux
should be those naturally appearing at future null infinity.

To be more concrete on this point,
let us consider a sphere $S$ at a finite distance belonging to a 
timelike hypersurface $\Gamma$; and which can be regarded as one of
a sequence of limiting spheres to future null infinity. Then in a numerical calculation
one constructs these spheres far enough and aims to have enough control on boundary 
conditions so that incoming radiation through $\Gamma$ is negligible. In this scenario,
the estimation of total momentum by using the limiting behavior of fields in the interior,
gives accurate information of the total momentum of a limiting sphere at future null infinity
after gauge issues are appropriately handled.

Last, we find it worthwhile to stress that the calculation
of momentum changes from the asymptotic fields at future null infinity 
does take into account all contributions in the interior of the spacetime,
including incoming radiation that ultimately might cross the horizon
of a final black hole.
The role of this radiation in the computed kick velocity is accounted by the measured 
variation of spatial momentum
and  black hole mass which changes as energy flows into the horizon. 
 Thus, the expression $v^i_{\mbox{kick}} = \Delta P^i / P^0$
 accounts for all relevant contributions.


\section{Acknowledgments}

This
work was supported in part by grants from NSF: PHY-0653375 and
PHY-0653369 to Louisiana State University, and
ANPCyT, CONICET and SeCyT-UNC.
L.L. wishes to thank the University of Cordoba and
the Aspen Center for Physics for hospitality
where parts of this work were completed.

\appendix
\section{}
The transformation between the dyad $\left\{o^A,\iota^A\right\}$ and $\left\{\tilde o^A,\tilde \iota^A\right\}$ induces a transformation on the corresponding null tetrads.
If we work with regular quantities at future null infinity we have,
\begin{eqnarray}
\hat\ell^a&=&\Omega^{-2}\ell^a,\\
\hat m^a&=&\Omega^{-1} m^a,\\
\hat n^a&=& n^a; 
\end{eqnarray}
and similar expression for its 'tilde' version. Using the relations (\ref{eq:regulariota}) and (\ref{eq:regularomicrom}), and having into account that the conformal factors are related as 
\begin{equation}\label{eq:conformalfactor}
\tilde\Omega=w \, \Omega;
\end{equation}
we find 
\begin{eqnarray}
\hat{\tilde{\ell}}^a&=&\frac{n^0(\alpha)}{w^2}\left(\hat\ell^a+\bar C\hat m^a+C\hat{\bar m}^a+C\bar C\hat n^a\right),\\
\hat{\tilde{m}}^a&=&\frac{e^{i\lambda}}{w}\left(\hat m^a+ C\hat n^a\right),\\
\hat{\tilde{n}}^a&=&\frac{1}{n^0(\alpha)}\hat n^a;
\end{eqnarray}
where in order to simplify the notation we have introduced
\begin{equation}
e^{i\lambda}\equiv\left({\frac{ \bar{\tilde\zeta}_{\bar\zeta} }{ \tilde\zeta_\zeta} }\right)^{\frac{1}{2}}.
\end{equation}
From this transformation of the regular null tetrad we can see that it is originated as a combination of 
transformations of types II and III.
The factor $n^0(\alpha)$ represents locally the action of a boost, 
similarly, $C$ the action of a null rotation around $\hat n$ and 
$e^{i\lambda}$ a spatial rotation. The factor $w$ comes as a result of both null
tetrads belonging to the same conformal metric class (see eq.(\ref{eq:conformalfactor})).

\section{}
The calculation of $\tilde\sigma_0$ is conveniently done referring to a 
`starred' tetrad
* which is defined in the following way.
Let us first define a * coordinate system such that
\begin{equation}
\ell = \frac{\partial}{\partial r^*};
\end{equation}
in other words, $r^*$ is an affine parameter of the null
geodesic vector defined by
\begin{equation}
\ell = du .
\end{equation}
We complete the coordinate system by
\begin{eqnarray}
u^* &=& u ,\\
r^* &=& \frac{r}{A}+ r_0 ,\\
\zeta^* &=& \zeta ,\\
\bar\zeta^* &=& \bar\zeta .
\end{eqnarray}
The * null tetrad related to this coordinate system can be given in
terms of the original tetrad by a transformation of the form
\begin{eqnarray}
\ell^* &=& \ell ,\\
m^* &=& \gamma \left( m + \Gamma \ell\right)  ,\\
n^* &=& n+\Gamma \bar m + \bar\Gamma m + \Gamma  \bar\Gamma \ell .
\end{eqnarray}
One can show that in our case $\gamma=1$ and
\begin{equation}
\Gamma = \frac{1}{A^2} \eth_V A .
\end{equation}

The null tetrad $(\ell^*,n^*,m^*,\bar m^*)$ have the characteristic
that is associated to a system $(u^*,r^*,\zeta^*,\bar\zeta^*)$, were
$r^*$ an affine radial coordinate. Then we can use the known
behavior of the spin coefficients in this tetrad-coordinate system
\cite{Moreschi87} in order to relate the leading order behavior of
this quantities with those obtained in the more general tetrad
$(l,n,m,\bar m)$.

In order to compute $\tilde\sigma_0$, we must know the leading order
behavior of $\tau$,$\beta$, $\epsilon'$, $\rho'$ and $\kappa'$.

The spin coefficients transform according to
\begin{equation}
\rho^* = \rho ,
\end{equation}
\begin{equation}
\sigma^* = \sigma ,
\end{equation}
\begin{equation}
\tau^* = \tau + \Gamma \rho
+\bar\Gamma \sigma ;
\end{equation}
which implies
\begin{equation}
\tau = \tau^*  - \Gamma \rho^*
-\bar\Gamma \sigma^* .
\end{equation}
From this one can deduce that the leading order behavior of the
original $\tau$ has a $1/r$ term, namely
\begin{equation}
\tau_{-1} = \frac{1}{A} \eth_V A .
\end{equation}

Similarly one has
\begin{equation}
\beta^* = \beta + \bar\Gamma \sigma + \Gamma \epsilon ;
\end{equation}
with
\begin{equation}
\epsilon^* = \epsilon ;
\end{equation}
but our choice of tetrad is such that
\begin{equation}
\epsilon = 0.
\end{equation}
Then in leading order the betas coincide, that is
\begin{equation}
\frac{\beta_0}{r} =
\frac{\beta^*_0}{r^*}
=A\frac{\beta^*_0}{r} .
\end{equation}
But one has
\begin{equation}
\beta^*_0 =
-\frac{V^*}{\sqrt{2}}\frac{\partial P^*_0}{\partial \zeta^*}
-\frac{1}{2V^*} \eth_{V^*} V^*
;
\end{equation}
and using that
\begin{equation}
V^* = \frac{V}{A} ;
\end{equation}
one obtains
\begin{equation}
\beta^*_0 =
\frac{\beta_{0V}}{A} + \frac{1}{2 A^2} \eth_{V} A
;
\end{equation}
with
\begin{equation}
\beta_{0V} =
-\frac{V}{\sqrt{2}}\frac{\partial P_0}{\partial \zeta}
-\frac{1}{2V} \eth_{V} V
.
\end{equation}
Then one has
\begin{equation}
\beta_0 =
\beta_{0V} + \frac{1}{2 A} \eth_{V} A
.
\end{equation}

Let us also note that
\begin{equation}
-\epsilon'^* =
-\epsilon' -\Gamma \beta' + \bar\Gamma (\tau+\beta)
+\bar\Gamma^2 \sigma + \Gamma \bar\Gamma \rho
;
\end{equation}
and
\begin{equation}
-\beta'^* = -\beta' + \bar\Gamma \rho .
\end{equation}
Using that in the first two orders in the expansion of
$\epsilon'$ one has
\begin{equation}
\epsilon' = \frac{1}{2} \frac{\partial U}{\partial r}
+O(\frac{1}{r}) ;
\end{equation}
one can deduce that
\begin{equation}
\hat n (\Omega C) + 2 \Omega C \epsilon' =
\Omega n^0(C) ;
\end{equation}
which was used above.

Finally one can see that the other spin coefficients needed in order
to compute $\sigma$ have the following behavior

\begin{equation}
\begin{split}
\rho'^*=\,&\rho+2\Gamma\bar\Gamma\epsilon-\Gamma\tau'+\bar\Gamma^2\sigma
-\bar\Gamma\left(\beta'-\beta\right)\\&+\Gamma\bar\Gamma^2\kappa+\eth\bar\Gamma+\Gamma
\ell(\bar\Gamma),
\end{split}
\end{equation}
and
\begin{equation}
\kappa'=\Omega \hat m(U);
\end{equation}
then one can see that they do not contribute to $\tilde\sigma_0$.

Collecting all these quantities in the expression for
$\tilde\sigma$, one has to leading order
\begin{equation}
\frac{\tilde \sigma_0}{\tilde r^2}
=
\frac{(w B)^2}{r^2}
\left({\frac{ \bar{\tilde\zeta}_{\bar\zeta} }{ \tilde\zeta_\zeta} }\right)^{\frac{1}{2}}
\left[
\sigma_0
-
\eth_V C
+ 2 \frac{C}{A} \, \eth_V A
- C \, n^0(C)
\right] ;
\end{equation}
which implies equation (\ref{eq:sigmatransformado}).


\end{document}